\documentclass[conference]{IEEEtran}

\usepackage[dvipsnames]{xcolor}
\usepackage{epsfig}
\usepackage{times}
\usepackage{float}
\usepackage{afterpage}
\usepackage{amsmath}
\usepackage{amstext}%
\usepackage{amssymb,bm}
\usepackage{latexsym}
\usepackage{color}
\usepackage{graphicx}
\usepackage{amsmath}
\usepackage{amsthm}
\usepackage{graphicx}
\usepackage[center]{caption}
\usepackage{pstricks}
\usepackage{caption}
\usepackage{subcaption} 
\usepackage{booktabs}
\usepackage{multicol}
\usepackage{lipsum}%
\usepackage[T1]{fontenc}
\usepackage{hyperref}
\usepackage{aecompl}
\usepackage{mathrsfs}

\long\def\comment#1{}

\newcommand{\av}{{\mathbf a}}
\newcommand{\bv}{{\mathbf b}}

\newcommand{\dv}{{\mathbf d}}
\newcommand{\ev}{{\mathbf e}}

\newcommand{\gv}{{\mathbf g}}

\newcommand{\pv}{{\mathbf p}}
\newcommand{\qv}{{\mathbf q}}

\newcommand{\xv}{{\mathbf x}}

\newcommand{\Fm}{{\mathbf F}}
\newcommand{\Gm}{{\mathbf G}}

\newcommand{\Bc}{{\mathcal B}}

\newcommand{\Gc}{{\mathcal G}}

\newcommand{\Ic}{{\mathcal I}}

\newcommand{\Qc}{{\mathcal Q}}

\newcommand{\Sc}{{\mathcal S}}
\newcommand{\Tc}{{\mathcal T}}

\newcommand{\Wc}{{\mathcal W}}

\newcommand{\qsf}{{\mathsf q}}
\newcommand{\rsf}{{\mathsf r}}

\newcommand{\Bsf}{{\mathsf B}}

\newcommand{\Ksf}{{\mathsf K}}
\newcommand{\Lsf}{{\mathsf L}}
\newcommand{\Msf}{{\mathsf M}}
\newcommand{\Nsf}{{\mathsf N}}

\newcommand{\Rsf}{{\mathsf R}}

\newcommand{\Tbp}{{\mathrm{T}}}

\theoremstyle{definition}
\newtheorem*{rem*}{Remark}

\theoremstyle{plain}
\newtheorem{thm}{Theorem}%

\newtheorem{rem}{Remark}

\providecommand{\definitionname}{Definition}

\let\svthefootnote\thefootnote
\newcommand\freefootnote[1]{%
  \let\thefootnote\relax%
  \footnotetext{#1}%
  \let\thefootnote\svthefootnote%
}

\title{Demand Privacy in Hotplug Caching Systems
}

\begin{document}

\author{
\IEEEauthorblockN{Yinbin Ma and Daniela Tuninetti}
\IEEEauthorblockA{University of Illinois Chicago, Chicago, IL 60607, USA \\ Email:\{yma52, danielat\}@uic.edu}
}

\allowdisplaybreaks
\maketitle
\IEEEpeerreviewmaketitle

\begin{abstract}
Coded caching, introduced by Maddah-Ali and Niesen (MAN), %
is a model where a server broadcasts multicast packets to users with a local cache that is leveraged so as to reduce the peak network communication load. 
The original MAN model does not consider missing demands (i.e., some users may not request a file) or privacy issues (i.e., decoding the multicast packets may expose the users' demands).
The former issue was captured by the
{\it hotplug model with offline users}, %
where the server starts sending multicast packets after having received a certain number of file requests.
The latter issue was addressed by devoting part of the cache to store
{\it privacy keys} 
to help users decode their requested file 
while remaining completely ignorant about the demands of the remaining users.
This paper investigates the problem of private demands against colluding users in the hotplug model with offline users.
Two achievable schemes are proposed based on Maximum Distance Separable (MDS) codes. 
They achieve lower subpacketization, and  
lower load in the small memory regime
compared to baseline schemes that trivially include demand privacy or offline users in known schemes.
\end{abstract}


\section{Introduction}
\label{sec:intro}
In a coded caching system a central server, with multiple files stored in it, connects to multiple users, each equipped with limited local cache space, through an error-free shared link. This model, introduced by Maddah-Ali and Niesen (MAN)~\cite{maddah2014fundamental}, aims to reduce the network peak load by leveraging the local caches.
Coded caching is attractive because it achieves impressive performance gains compared to uncoded systems~\cite{maddah2014fundamental}. 
The system works in two phases:
in the {\it cache placement phase}, the server populates the caches without knowing the future users' demands; 
in the {\it delivery phase}, the server broadcasts coded multicast messages to the users to meet their demands. 
For single file retrieval and under the constraint of {\it uncoded placement}
\footnote{File bits are directly copied into the caches without coding.} 
a converse bound was derived in~\cite{wan2020index} and proved to match the MAN achievable scheme~\cite{maddah2014fundamental} when there are more files than users. Later, Yu {\it et al.} proposed an achievable scheme (referred to as the YMA scheme) that matches the converse bound in~\cite{wan2020index} also when there are less files than users.  
Coded placement, as in~\cite{amiri2016fundamental,zhang2018fundamental}, is proved to reduce the YAM load by no more than a factor of two~\cite{yu2018characterizing}. Coded placement schemes use linear codes, such as rank metric codes or Maximum Distance Separable (MDS) codes~\cite{zhang2018fundamental}. %
Wan {\it et al.} extended the single file retrieval model in~\cite{maddah2014fundamental} so as to include Scalar Linear Function Retrieval (SLFR)~\cite{wan2020fundamental},  where users can demand a linear function of the files. 
By generalizing the YMA scheme %
to any finite field, Wan {\it et al.} proved that SLFR attains the same load in single file retrieval in~\cite{wan2020fundamental}.
A general SLFR construction can be found in~\cite{ma2021general}.

In practical scenarios some users may fail to send their demand to the server due to technical failures. Since all users' demands are required to generate MAN multicast messages, missing demands cause the system to halt as the server cannot enter the delivery phase. %
Relevant work addressing missing demands include the {\it asynchronous demands}~\cite{maddah2014fundamental} and the {\it hotplug with offline users}~\cite{ma2021coded}
\footnote{Hotplug is a computer system term that refers to a device that can be added or removed from the running system without having to restart the system.} settings. 
In the hotplug model, the server starts transmitting as soon as a certain number of demands is received, regardless of who the offline users are, where the number of active users (but not their identity) is assumed to be known at the time of placement.
This latter assumption has been relaxed in~\cite{liao2021fundamental} by considering average load (as opposed to peak load) and by allowing the server to fail delivery with a certain `outage' probability.
Another practical concern in coded caching is preserving {\it demand privacy}. 
Private Information Retrieval (PIR)~\cite{chor1995private} addresses the demand privacy issue against multiple servers in a single user system. Here we are interested in protecting demand privacy of each user against the remaining users~\cite{wan2020codedprivate, gurjarpadhye2022fundamental}.
Most coded caching schemes, including the YMA scheme, users need to know all the demands in order to decode multicast messages. %
Demand private schemes include the {\it virtual users} scheme in~\cite{engelmann2017content,kamath2019demand} (in which the server pretends to serve many more users than actually present in the system) 
and the {\it privacy keys} scheme in~\cite{yan2021fundamental} (where keys are stored and used to confuse users so as to prevent them from gaining any knowledge about the demands). Both schemes are demand private even against colluding users~\cite{yan2021fundamental} or in D2D settings~\cite{yapar2019optimality}.

{\bf This paper investigates the problem of SLFR demand privacy against colluding users in the hotplug model with offline users.}
Two novel achievable schemes with MDS coding in the placement phase are proposed. 
They are shown to achieve lower subpacketization, and lower load in the small memory regime, compared to baseline schemes that trivially include demand privacy or offline users in known schemes.
In Appendix, they are numerically compared to known converse bounds for systems without privacy or without offline users.

The rest of the paper is organized as follows.
Section~\ref{sec:formulation} states the investigated problem and summarizes related known results.
Section~\ref{sec:main} summarizes the main results of this paper.
Section~\ref{sec:prooflayerscheme} proves the main theorems. %
Section~\ref{sec:conclusion} concludes the paper.
Some proofs and experiments can be found in Appendix.

\section{Problem Formulation and Known Results}
\label{sec:formulation}

\subsection{Notation Convention}
We adopt the following notation convention.
Calligraphic symbols denote sets, bold lowercase symbols 
vectors, bold uppercase symbols matrices, and sans-serif symbols system parameters. $|\cdot|$ is the cardinality of a set or the length of a vector. For integers $a$ and $b$, $\binom{a}{b}$ is the binomial coefficient, or 0 if $a \geq b \geq 0$ does not hold. For an integer $b$, we let $[b] := \{1, \ldots, b\}$. For sets $\Sc$ and $\Qc$, we let $\Sc \setminus \Qc := \{k: k \in \Sc, k \notin \Qc\}$. For a collection $\{Z_1, \ldots Z_n\}$ and a index set $\Sc \subseteq [n]$, we let $Z_\Sc := \{Z_i: i \in \Sc\}$. For a set $\Gc$ and an integer $t$, we let $\Omega_{\Gc}^{t} := \{ \Tc \subseteq \Gc : |\Tc| = t\}$. When necessary, we denote $\Omega_{[n]}^t = \{\Tc_1, \ldots, \Tc_{\binom{n}{t}} \}$, where $\Tc_i$ be the $i$-th subset in $\Omega_{[n]}^t$ in lexicographical order. For example,  $\Omega_{[3]}^2 = \{\Tc_1,\Tc_2,\Tc_3\}$ with $\Tc_1 = \{1,2\}$, $\Tc_2 = \{1,3\}$, $\Tc_3 = \{2,3\}$.

\subsection{Problem Formulation}\label{sec:model}
The $(\Ksf, \Ksf^\prime, \Nsf)$ hotplug system~\cite{ma2021coded} includes a server, $\Ksf$ users, $\Nsf$ files, and $\Ksf^\prime \leq \Ksf$. Each file has $\Bsf$ symbols which are uniformly and independently distributed over $\mathbb{F}_\qsf$, where $\qsf$ is a prime-power number. Files are denoted by $F_i \in \mathbb{F}_{\qsf}^{\Bsf}, i\in[\Nsf]$. All users are connected to the server via an error-free shared link. Each user can store $\Msf \Bsf$ symbols in its local cache; we refer to $\Msf$ as the {\it memory size}. 
In the placement phase, the server pushes content in the caches without knowledge of future demands. 
The cache content of user~$j\in[\Ksf]$ is denoted by $Z_j \in \mathbb{F}_{\qsf}^{\Msf\Bsf}$. 
In the delivery phase, $\Ksf^\prime$ users become active, and are indexed by $\Ic \in \Omega_{[\Ksf]}^{\Ksf^\prime}$; active user~$j \in \Ic$ demands $\dv_j = (d_{j,1}, \ldots d_{j,\Nsf}) \in \mathbb{F}_{\qsf}^{\Nsf}$, meaning that it is interested in retrieving the scalar linear combination of files denoted by
\begin{align}
\langle  \dv_j, F_{[\Nsf]} \rangle := \sum_{n \in [N]} d_{j,n} F_n \in \mathbb{F}_{\qsf}^{\Bsf}.
\label{eq:SLFRdemand}
\end{align} 
Note the abuse of notation in~\eqref{eq:SLFRdemand}: we use the inner product notation, however here the $d_{j,n}$'s are scalar but $F_n$'s are the files seen as vectors of length $\Bsf$.
We assume the demands are independent across users. 
The server collects the demands of the active users in the matrix $\dv_{\Ic} \in \mathbb{F}_{\qsf}^{\Ksf^\prime \times \Nsf}$, and transmits a message $X\in \mathbb{F}_{\qsf}^{\Rsf\Bsf}$, which is a function of $\dv_{\Ic}$; %
we refer to $\Rsf$ as the {\it load}.
Each user must decode its demanded SLFR based on its local cached content and the delivery signal $X$.

As in past work on coded caching with privacy~\cite{wan2020codedprivate,yan2021fundamental} or security~\cite{yan2022robust}, we assume that each user~$j \in [\Ksf]$ has available some {\it local randomness} represented by the random variable (RV) $\tau_j$. We assume that $\tau_j$ is only known to the server and to user~$j \in [\Ksf]$, and is independent of other RVs. The local randomness RVs help to generate the cache contents and the transmit signal $X$ and may be stored if needed for decoding.

Mathematically, the constraints in our problem are
\begin{align}
    & H(F_{[\Nsf]}) = H(F_i) + \ldots + H(F_\Nsf) = \Nsf \Bsf \log(\qsf), %
     \label{eq:equalandindptfiles} \\
    & H(Z_j | F_{[\Nsf]}, \tau_j) = 0, \ \forall j \in [\Ksf],
    \label{eq:cacheconstraint}\\
    & H(X | \dv_\Ic, F_{[\Nsf]}, \tau_{[\Ksf]} ) = 0, \ \forall \Ic \in \Omega_{[\Ksf]}^{\Ksf^\prime},
    \label{eq:messageconstraint}\\
    & H(\langle  \dv_j, F_{[\Nsf]} \rangle  | X, \dv_j, Z_j) = 0, \ \forall j \in \Ic, 
    \label{eq:correctnessconstraint}\\
    & I(\dv_{\Ic\setminus\Bc}; X, \dv_\Bc, Z_\Bc \vert F_{[\Nsf]}) = 0, \ \forall \Bc \subseteq \Ic, \Bc \not=\emptyset, %
    \label{eq:privacyconstraint}
\end{align}
where
\eqref{eq:equalandindptfiles} is because the $\Nsf$ files are composed of $\Bsf$ i.i.d. symbols over $\mathbb{F}_\qsf$, 
\eqref{eq:cacheconstraint} is the cache placement,
\eqref{eq:messageconstraint} is the delivery signal,
\eqref{eq:correctnessconstraint} is the condition for correct decoding, and
\eqref{eq:privacyconstraint} is the privacy condition as in~\cite{yan2021fundamental}. 

Our goal is to characterize the {\it worst-case load} defined as
\begin{align}
\Rsf&^{\star}(\Msf)
 = \limsup_{\Bsf, \qsf } %
 \ {\min_{X,Z_{[\Ksf]}}}
 \ \max_{\Ic,\dv_{\Ic}} \{\Rsf: 
\textrm{\textrm{\small conditions in~\eqref{eq:equalandindptfiles}-\eqref{eq:privacyconstraint}}}\notag\\&\quad
\textrm{are satisfied with memory size $\Msf$}\}, \ \Msf \in [0, \Nsf].
\end{align}
We denote by $\Lsf \in \mathbb{Z}_{+}$ the {\it subpacketization level}, which is the smallest $\Bsf$ needed for a given achievable scheme, which measures the implementation complexity of the scheme.

\subsection{Privacy Keys Scheme for the Classic SLFR Model}\label{sec:YT}
Yan {\it et al.} in~\cite{yan2021fundamental} introduced an achievable scheme that preserves demand privacy for the classic (not hotplug) model, which is equivalent to $\Ksf^\prime=\Ksf$ in the model in Section~\ref{sec:model}. The privacy keys scheme in~\cite{yan2021fundamental} works as follows. 

\paragraph{File Partition} 
Fix $t \in [0 : \Ksf]$. We partition each file into equal-length subfiles as follows. With a Matlab-like notation and assuming $\Lsf_t=\binom{\Ksf}{t}$ divides $\Bsf$, we define
\begin{align}
    F_i &= (F_{i,\Wc} : \Wc \in \Omega_{[\Ksf]}^t) \in \mathbb{F}_\qsf^{\Bsf}, \ \forall i \in [\Nsf],
    \\
    F_{i,\Wc} &:= F_i(\Wc) \in \mathbb{F}_\qsf^{\Bsf/\binom{\Ksf}{t}}, \ \forall \Wc \in \Omega_{[\Ksf]}^t. 
    \label{eq:filepartition}
\end{align}

Given vectors $\av \in \mathbb{F}_\qsf^{\Nsf}$ %
and $\bv \in \mathbb{F}_\qsf^{\binom{\Ksf}{t}}$, %
we denote
the ``bilinear product of the subfiles'' as
\begin{align}
    \Tbp(\av, \bv) := %
    \sum_{n \in [\Nsf]} \sum_{j\in\left[\binom{\Ksf}{t}\right]} a_n \, F_{n, \Wc_j} \, b_j \in \mathbb{F}_\qsf^{\Bsf/\binom{\Ksf}{t}}. 
    \label{eq:bilinearproduct}
\end{align}
Let $\ev_{i}$ be the vector with only one non-zero component equal to 1 in position $i$.
We can write $F_{i,\Wc_j}$ in~\eqref{eq:filepartition} as $F_{i,\Wc_j} = \Tbp(\ev_{i}, \ev_{j})$;  with a slight abuse of notation we shall sometimes write $F_{i,\Wc} = \Tbp(\ev_{i}, \ev_{\Wc})$, for $i \in [\Nsf], \ \Wc \in \Omega_{[\Ksf]}^t$.

\paragraph{Cache Placement} For each user~$j \in [\Ksf]$, the server i.i.d uniformly at random generates the random vector $\pv_j \in \mathbb{F}_\qsf^{\Nsf}$ and populates the cache of user~$j$ as
\begin{subequations}
\begin{align}
    Z_j &= \left\{\Tbp(\ev_i, \ev_{\{j\}\cup\Qc}): i \in [\Nsf], \Qc \in \Omega_{[\Ksf] \setminus \{j\}}^{t-1}\right\} 
    \label{eq:place:man}
    \\ &\cup \ \left\{\Tbp(\pv_j, \ev_{\Wc}): \Wc \in \Omega_{[\Ksf]\setminus\{j\}}^{t}\right\}, 
    \label{eq:place:pk}
\end{align}
\end{subequations}
where~\eqref{eq:place:man} is the classical uncoded MAN placement and~\eqref{eq:place:pk} are the privacy keys. The needed memory size is
\begin{align}
    \Msf^\text{\rm(PK)}_t = \Nsf \frac{\binom{\Ksf-1}{t-1}}{\binom{\Ksf}{t}} + \frac{\binom{\Ksf-1}{t}}{\binom{\Ksf}{t}} = 1 + \frac{t}{\Ksf}(\Nsf-1)\geq 1.
    \label{eq:YT:memory}
\end{align}
Note that $\pv_j$ is the local randomness for user~$j\in[\Ksf]$, which is not stored directly in the cache.

\paragraph{Delivery Phase}
For user~$j \in [\Ksf]$ with demand $\dv_j\in \mathbb{F}_\qsf^{\Nsf}$, %
let $\qv_j = \pv_j + \dv_j\in \mathbb{F}_\qsf^{\Nsf}$.
The server transmits 
\begin{subequations}
\begin{align}
   X^\text{\rm(PK)} &= \{X^\text{\rm(PK)}_\Sc: \Sc \in \Omega_{[\Ksf]}^{t+1} \} \cup \{ \qv_{[\Ksf]} \}, 
   \label{eq:YTmessages:X}
\end{align}
where we can write
\begin{align}
X^\text{\rm(PK)}_\Sc 
    &= \sum_{j \in \Sc} \Tbp(\qv_j, \ev_{\Sc \setminus \{j\}})
    \\
    &= \underbrace{ \Tbp(\dv_v, \ev_{\Sc \setminus \{v\}}) }_{\text{demanded by user~$v$}}
     + \underbrace{ \Tbp(\pv_v, \ev_{\Sc \setminus \{v\}}) }_{\text{from cache $Z_v$ in~\eqref{eq:place:pk}}}
    \label{eq:YTmessages:S1}
    \\ 
    &+
    \sum_{j \in \Sc \setminus \{v\}} \Tbp(\qv_j, \ev_{\Sc \setminus \{j\}}) , \ \forall  v\in \Sc,
    \label{eq:YTmessages:S2}
\end{align}
and, from the cache content $Z_v$ in~\eqref{eq:place:man} and $\qv_{[\Ksf]}$ in~\eqref{eq:YTmessages:X}, we can locally compute all terms in~\eqref{eq:YTmessages:S2} as
\begin{align}
    \Tbp(\qv_j, \ev_{\Sc \setminus \{j\}}) = \sum_{n\in[\Nsf]} q_{j,n}\Tbp(\ev_n, \ev_{\Sc \setminus \{j\}}).
    \label{eq:YTmessages:S2evallocally}
\end{align}
\label{eq:YTmessages}
\end{subequations}
Therefore, user~$v\in\Sc$ can retrieve $\Tbp(\dv_v, \ev_{\Sc \setminus \{v\}})$ from $X^\text{\rm(PK)}_\Sc$.
If all $\Sc \in \Omega_{[\Ksf]}^{t+1}$ are sent, clearly every active user recovers its demanded SLFR.
Some of the $X^\text{\rm(PK)}_\Sc$'s are linearly dependent on the others 
and can be removed as explained in~\cite{wan2021optimal};
thus, in the limit for $\Nsf\Ksf/\Bsf\to0$ so that sending $\qv_{[\Nsf]}$ in~\eqref{eq:YTmessages:X} uses negligible resources, the following load is achievable
\begin{align}
    (\Msf^\text{\rm(PK)}_t, \Rsf^\text{\rm(PK)}_t) = 
    \left( 1 + \frac{t}{\Ksf}(\Nsf-1),
    \frac{\binom{\Ksf}{t+1} - \binom{\Ksf - \rsf}{t+1}}{\binom{\Ksf}{t}} \right),
    \label{eq:YT:load}
\end{align}
with subpacketization $\Lsf_t=\binom{\Ksf}{t}$ and with $\rsf = \min(\Ksf, \Nsf)$. %

Therefore, the lower convex envelope of the points in~\eqref{eq:YT:load} for $t \in [0 : \Ksf]$, together with the trivial point $(\Msf,\Rsf)=(0,\Nsf)$, is achievable~\cite[Theorem~2]{yan2021fundamental}. It is interesting to note that the load in~\eqref{eq:YT:load} is the same as for single file retrieval without privacy~\cite{yu2017exact} for all $t \in [0 : \Ksf]$, where the price of privacy is in cache size $\Msf^\text{\rm(PK)}_t$ in~\eqref{eq:YT:memory} needed to achieve this load compared to the classical model (i.e.,  $\Msf^\text{(MAN)}_t=\frac{t}{\Ksf}\Nsf \leq \Msf^\text{\rm(PK)}_t$).

\section{Main Results}\label{sec:main}

\begin{thm}[Baseline Scheme~1: Extension of the virtual users scheme~\cite{kamath2019demand}] \label{thm:basevirtualscheme}
    For a $(\Ksf, \Ksf^\prime, \Nsf)$ hotplug model with constraints~\eqref{eq:equalandindptfiles}-\eqref{eq:privacyconstraint}, the lower convex envelope of $(\Msf,\Rsf)=(0,\Nsf)$ and the following points is achievable, for $\widetilde{\Nsf} = \frac{\qsf^\Nsf-1}{\qsf-1}$
    \begin{align}
        ( \Msf^\text{\rm(BL1)}_t, \Rsf^\text{\rm(BL1)}_t ) 
        =\left( \frac{t}{\widetilde{\Nsf} \Ksf}\Nsf, \frac{\binom{\Ksf\widetilde{\Nsf}}{t+1} - \binom{\Ksf\widetilde{\Nsf} - \widetilde{\Nsf}}{t+1}}{\binom{\Ksf\widetilde{\Nsf}}{t}}  \right),  
    \label{eq:performancevirt}
    \end{align}
    with subpacketization level $\binom{\Ksf\widetilde{\Nsf}}{t}$, for all $t \in [0 : \Ksf \widetilde{\Nsf}]$. 
\end{thm}

\begin{thm}[Baseline Scheme~2: Extension of the privacy keys scheme~\cite{yan2021fundamental}] \label{thm:baseytscheme}
    For a $(\Ksf, \Ksf^\prime, \Nsf)$ hotplug model with constraints~\eqref{eq:equalandindptfiles}-\eqref{eq:privacyconstraint}, the lower convex envelope of $(\Msf,\Rsf)=(0,\Nsf)$ and the following points is achievable with $\rsf^\prime = \min(\Ksf^\prime, \Nsf)$
    \begin{align}
        ( \Msf^\text{\rm(PK)}_t, \Rsf^\text{\rm(BL2)}_t ) = \left( 1+\frac{t}{\Ksf}(\Nsf -1),%
         \frac{\binom{\Ksf}{t+1} - \binom{\Ksf - \rsf^\prime}{t+1}}{\binom{\Ksf}{t}}  \right),
    \label{eq:performanceyt}
    \end{align}
    with subpacketization level $\binom{\Ksf}{t}$, for all $t \in [0 : \Ksf]$. 
\end{thm}

\begin{rem} \rm
    Theorems~\ref{thm:basevirtualscheme} and~\ref{thm:baseytscheme} 
    extend known schemes from the classic model to the hotplug model. The idea is that, even if the demands of the offline users are not known, the server can assign a fictitious demand to those offline users. 
    The load-memory tradeoff in Theorem~\ref{thm:basevirtualscheme} does not depend on the number of active users $\Ksf^\prime$, while it does in Theorem~\ref{thm:baseytscheme}. In Theorem~\ref{thm:baseytscheme}, %
    $\Rsf^\text{\rm(BL2)}_t \leq \Rsf^\text{\rm(PK)}_t$ for all $t \in [0 : \Ksf]$, because the number of distinct demands is at most $\rsf^\prime$ which is no larger than $\rsf$ in~\eqref{eq:YT:load}.
    Note that $\Rsf^\text{\rm(BL1)}_t$ depends on $\widetilde{\Nsf}$, the number of distinct non-zero linear combinations that can be demanded.
    Finally we note that Theorem~\ref{thm:basevirtualscheme} achieves the best performance for some system parameters, but its subpacketization level is huge and unaffordable for a practical system.
$\hfill\square$\end{rem}

\begin{thm}[New Scheme~1] \label{thm:layerscheme}
    For a $(\Ksf, \Ksf^\prime, \Nsf)$ hotplug model with constraints~\eqref{eq:equalandindptfiles}-\eqref{eq:privacyconstraint}, the lower convex envelope of $(\Msf,\Rsf)=(\Nsf,0)$ and the following points  is achievable
    \begin{multline}
        ( \Msf^\text{\rm(new)}_t, \Rsf^\text{\rm(new)}_t ) = \\ \left( 
        \frac{\Nsf \binom{\Ksf-1}{t-1}+ \left[\binom{\Ksf^\prime}{t}- \binom{\Ksf-1}{t-1}\right]^+}{\binom{\Ksf^\prime}{t}}, %
        \frac{\binom{\Ksf^\prime}{t+1} - \binom{\Ksf^\prime - \rsf^\prime}{t+1}}{\binom{\Ksf^\prime}{t}}  \right),
    \label{eq:performancelayer}
    \end{multline}
    where $t \in [0 : \Ksf^\prime]$, %
    with subpacketization level $\binom{\Ksf^\prime}{t}$.
\end{thm}

\begin{rem} \rm
    Theorem~\ref{thm:layerscheme} combines ideas from the hotplug achievable scheme in~\cite{ma2021coded} and the privacy keys scheme in Section~\ref{sec:YT}. We apply an MDS code of rate ${\binom{\Ksf^\prime}{t}}/{\binom{\Ksf}{t}}$ %
    to the subfiles, after file partition with subpacketization level $\binom{\Ksf^\prime}{t}$, to get a load that is the same as if the system only had $\Ksf^\prime$ overall users (i.e., load as in~\eqref{eq:performanceyt} with $\Ksf$ replaced by $\Ksf^\prime$) which comes at a  price of storing MDS-coded privacy keys, i.e., 
\begin{align}
\Msf^\text{\rm(new)}_t
\geq 1+\frac{t}{\Ksf^\prime} \frac{\binom{\Ksf-1}{t-1}}{\binom{\Ksf^\prime-1}{t-1}} ( \Nsf - 1)
\geq 1+\frac{t}{\Ksf^\prime} ( \Nsf - 1).
\end{align}
Moreover, %
Theorem~\ref{thm:layerscheme} has a lower subpacketization level than Theorem~\ref{thm:baseytscheme} while maintaining a competitive performance.
$\hfill\square$
\end{rem}

If the server would have the ForeSight (FS) to know which subsets of users will be active during the delivery phase at the time of placement, we could hope to achieve the memory-load tradeoff for demand private caching with $\Ksf^\prime$ users, namely
    \begin{align}
        ( \Msf^\text{\rm(FS)}_t, \Rsf^\text{\rm(FS)}_t ) = \left( 1+\frac{t}{\Ksf^\prime}(\Nsf -1),
         \frac{\binom{\Ksf^\prime}{t+1} - \binom{\Ksf^\prime - \rsf^\prime}{t+1}}{\binom{\Ksf^\prime}{t}}  \right),
    \label{eq:foresight}
    \end{align}
    for all $t \in [0 : \Ksf^\prime]$. 
Notice that $( \Msf^\text{\rm(FS)}_t, \Rsf^\text{\rm(FS)}_t )$ is the same as $( \Msf^\text{\rm(PK)}_t, \Rsf^\text{\rm(PK)}_t )$ after replacing $\Ksf$ with $\Ksf^\prime$.
Theorem~\ref{thm:layerscheme} shows (proved in Appendix) that we achieve this ``foresight'' performance in the small cache regime (i.e., $\Msf\in[0,1+\frac{\Nsf-1}{\Ksf^\prime}]$). 
Theorem~\ref{thm:barescheme} next shows that this ``foresight'' performance is also achievable in the large cache size regime (i.e., $\Msf\in[\Nsf - \frac{\Nsf-1}{\Ksf^\prime},\Nsf]$).

\begin{thm}[New Scheme~2] 
\label{thm:barescheme}
For a $(\Ksf, \Ksf^\prime, \Nsf)$ hotplug model with constraints~\eqref{eq:equalandindptfiles}-\eqref{eq:privacyconstraint}, the segment connecting $(\Msf^\text{\rm(FS)}_{\Ksf^\prime-1}, \Rsf^\text{\rm(FS)}_{\Ksf^\prime-1}) = (\Nsf - (\Nsf-1)/\Ksf^\prime, 1/\Ksf^\prime)$ and $(\Msf^\text{\rm(FS)}_{\Ksf^\prime}, \Rsf^\text{\rm(FS)}_{\Ksf^\prime}) = (\Nsf,0)$ is achievable. %
\end{thm}
We provide the proof of Theorem~\ref{thm:barescheme} in Appendix.

\begin{rem}[Converse] \label{rem:converse}
\rm
In this paper we focus on achievable schemes. A trivial converse bound can be obtained by considering the converse bound in~\cite[Theorem~3]{yan2021fundamental} (for demand privacy with colluding users and for single file retrieval) with $\Ksf$ replaced by $\Ksf^\prime$. The reasoning is that we cannot do better than knowing at the time of placement which $\Ksf^\prime$ users will be active, and restricting them to request a single file as opposed to a scalar linear combination of files. 
The lower bound in~\cite[Theorem~3]{yan2021fundamental}
is known to be `good' in the low memory regime~\cite{yan2021fundamental}, i.e., note $\Rsf^\star(0) \geq \Nsf.$
In the large memory regime we use the converse bound in~\cite[Theorem~2]{yu2018characterizing} (without any privacy) again with $\Ksf$ replaced by $\Ksf^\prime$.
In our plots we use the largest of these two lower bounds.
$\hfill\square$\end{rem}

\section{Proof of Theorem~\ref{thm:layerscheme}}\label{sec:prooflayerscheme}

\paragraph{File Partition} 
Fix $t \in [\Ksf^\prime]$ and partition the files into $\Lsf_t=\binom{\Ksf^\prime}{t}$ equal-length subfiles. 
Then, code the subfiles of each file with the MDS generator matrix $\Gm$, where any $\binom{\Ksf^\prime}{t}$ columns of $\Gm$ are linearly independent:
for all $n\in[\Nsf]$
\begin{align}
    &C_{n,\Tc_\ell} 
    := \sum_{j\in\left[\binom{\Ksf^\prime}{t}\right]}  F_{n, \Wc_j} \ g_{\ell,j}
    = \Tbp(\ev_n, \gv_{\ell}) \in \mathbb{F}_\qsf^{\Bsf/\binom{\Ksf^\prime}{t}},
    \label{eq: mds subfiles}
    \\
    &\gv_{\ell} 
    := (g_{\ell,1}, \ldots, g_{\ell,\binom{\Ksf^\prime}{t}})^T \in \mathbb{F}_\qsf^{\binom{\Ksf^\prime}{t}}, 
    \ \forall \ell\in\left[\binom{\Ksf}{t}\right], 
    \\
    &\Gm
    :=[\gv_{1}; \cdots; \gv_{\binom{\Ksf}{t}}]\in \mathbb{F}_\qsf^{\binom{\Ksf^\prime}{t} \times \binom{\Ksf}{t}}.
\end{align}
Note the slight abuse of notation in~\eqref{eq: mds subfiles}: the definition of $T(\cdot, \cdot)$ in~\eqref{eq:bilinearproduct} assumes subpacketization level $\binom{\Ksf}{t}$, while here the subpacketization level is $\binom{\Ksf^\prime}{t}$; we decided not to introduce a new notation in~\eqref{eq: mds subfiles} as we essentially use $T(\cdot, \cdot)$ to denote a linear combination of subfiles whose subpacketization level is clear from the context.

\paragraph{Cache Placement}
For each user~$j \in [\Ksf]$, the server generates an i.i.d uniformly random vector $\pv_j \in \mathbb{F}_\qsf^\Nsf$ and  places it into $Z_j$, together with the MAN-like content 
\begin{align}
\{\Tbp(\ev_n, \gv_\Tc): n \in [\Nsf], \Tc \in \Omega_{[\Ksf]}^{t}, j \in \Tc\}. 
\label{eq:LMDC:MAN-like}
\end{align}
To guarantee privacy, we also need to store (inspired by the PK scheme in~\eqref{eq:place:pk}) the privacy keys $\{\Tbp(\pv_j, \gv_\Wc) : \Wc \in \Omega_{[\Ksf]\setminus\{j\}}^{t} \}$ in the cache of user~$j \in [\Ksf]$. It would therefore seem that we need to store $\Nsf\binom{\Ksf-1}{t-1} + \binom{\Ksf-1}{t}$ subfiles as in the PK scheme. The privacy keys are however MDS coded, so we only need to store $\binom{\Ksf^\prime}{t}$ of them to recover all possible $\binom{\Ksf}{t}$. Now, from the MAN-like content in~\eqref{eq:LMDC:MAN-like} and the local randomness $\pv_j$, we can locally compute (similarly to~\eqref{eq:YTmessages:S2evallocally}) %
\begin{align}
\Tbp(\pv_j, \gv_\Tc) = \sum_{n\in[\Nsf]} p_{j, n}\Tbp(\ev_n, \gv_\Tc) : \Tc \in \Omega_{[\Ksf]}^{t}, j \in \Tc,
\label{eq:LMDC:from MAN-like}
\end{align}
this implies that if $\binom{\Ksf-1}{t-1} \geq \binom{\Ksf^\prime}{t}$ we are good; otherwise we need to store additional $\binom{\Ksf^\prime}{t} - \binom{\Ksf-1}{t-1}$ privacy keys that are linearly independent of those in~\eqref{eq:LMDC:from MAN-like}.
\begin{subequations}
The memory size is thus
\begin{align}
    \Msf^\text{\rm(new)}_t 
    &=\begin{cases} 
    \Nsf \frac{\binom{\Ksf-1}{t-1}}{\binom{\Ksf^\prime}{t}} &  \binom{\Ksf-1}{t-1} \geq \binom{\Ksf^\prime}{t} \\
    1+\frac{t}{\Ksf^\prime}\frac{\binom{\Ksf-1}{t-1}}{\binom{\Ksf^\prime-1}{t-1}}(\Nsf-1) &  \binom{\Ksf-1}{t-1} < \binom{\Ksf^\prime}{t} 
    \end{cases}.
\end{align}
\label{eq:LMDC:M}
\end{subequations}
Let $\eta_t := \left[\binom{\Ksf^\prime}{t}- \binom{\Ksf-1}{t-1}\right]^+$.
The cache  for user~$j \in [\Ksf]$ is 
\begin{subequations}
\begin{align}
Z_j &=  \left\{\Tbp(\ev_i, \gv_{\{j\}\cup\Qc}): i \in [\Nsf], \Qc \in \Omega_{[\Ksf] \setminus \{j\}}^{t-1}\right\}  
\label{eq:LMDC:ZjMAN}
\\ & \cup \ \left\{\Tbp(\pv_j, \xi_{j,\ell}): \ell \in \left[\eta_t\right]\right\} \cup \{\pv_j\},
\label{eq:LMDC:ZjPK}
\end{align}
\label{eq:LMDC:Zjall}
\end{subequations}
where $\{\xi_{j,\ell}: \ell \in [\eta_t]\}$ is a subset of $\{\gv_{\Tc}: \Tc \in \Omega_{[\Ksf]\setminus \{j\}}^{t}\}$ that is linearly independent w.r.t. $\{\gv_{\{j\}\cup\Qc}: \Qc \in \Omega_{[\Ksf] \setminus \{j\}}^{t-1}\}$.

\paragraph{Delivery Phase}
Given the active users indexed by $\Ic \in \Omega_{[\Ksf]}^{\Ksf^\prime}$ %
where user~$j \in \Ic$ demands $\dv_j$, %
we let $\qv_j = \pv_j + \dv_j$ as in the PK scheme. 
\begin{subequations}
The server sends
\begin{align}
  X &=  \{X_\Sc: \Sc \in \Omega_{\Ic}^{t+1} \} \cup \{ \qv_{\Ic}, \Ic \}, \ \text{where}
\label{eq:LayerMDSmessages:X}
  \\ 
  X_\Sc &= \sum_{j \in \Sc} \Tbp(\qv_j, \gv_{\Sc \setminus \{j\}}) \\
&= \underbrace{ \Tbp(\dv_v, \gv_{\Sc \setminus \{v\}}) }_{\text{demanded by user~$v$}}
     + \underbrace{ \Tbp(\pv_v, \gv_{\Sc \setminus \{v\}}) }_{\text{from cache $Z_v$ in~\eqref{eq:LMDC:ZjPK}}}
\\
    &+ \sum_{j \in \Sc \setminus \{v\}} \underbrace{\Tbp(\qv_j, \gv_{\Sc \setminus \{j\}})}_{\text{from cache $Z_v$ in~\eqref{eq:LMDC:ZjMAN} and $\qv_{\Ic}$ in~\eqref{eq:LayerMDSmessages:X}}}, \ \forall  v\in \Sc.
\end{align}
\label{eq:LayerMDSmessages}
\end{subequations}
As in the PK scheme, some multicast messages in~\eqref{eq:LayerMDSmessages:X} need not be sent (similarly to the PK scheme), %
resulting in the load
$
    \Rsf^\text{\rm(new)}_t = \bigl[\binom{\Ksf^\prime}{t+1} - \binom{\Ksf^\prime - \rsf^\prime}{t+1}\bigr] /\binom{\Ksf^\prime}{t}.
$
Note, for $t=0$ a similar idea can be used to achieve $(\Msf^\text{\rm(new)}_0,\Rsf^\text{\rm(new)}_0)=(1,\rsf^\prime)$, see also the example next.

\paragraph{Proof of Privacy}
We prove that our scheme satisfies the privacy constraint in~\eqref{eq:privacyconstraint}. Fix $\Bc \subseteq \Ic$, then
\begin{align*}
    & I(\dv_{\Ic \setminus \Bc};X, \dv_{\Bc}, Z_{\Bc} \vert F_{[\Nsf]}) \\
    & \stackrel{\rm{(a)}}{=} I(\dv_{\Ic \setminus \Bc}; \qv_{\Ic}, \dv_\Bc, Z_\Bc \vert F_{[\Nsf]}) %
    \\ &\stackrel{\rm{(b)}}{=} I(\dv_{\Ic \setminus \Bc}; \qv_{\Ic}, \dv_\Bc, \pv_\Bc \vert F_{[\Nsf]}) %
    \\ &\stackrel{\rm{(c)}}{=} I(\dv_{\Ic \setminus \Bc}; \qv_{\Ic \setminus \Bc}, \dv_\Bc, \pv_\Bc \vert F_{[\Nsf]}) %
    \\ &= I(\dv_{\Ic \setminus \Bc}; \dv_\Bc, \pv_\Bc \vert F_{[\Nsf]}) 
        + I(\dv_{\Ic \setminus \Bc}; \qv_{\Ic \setminus \Bc} \vert \dv_\Bc, \pv_\Bc, F_{[\Nsf]}) %
    \\ &\stackrel{\rm{(d)}}{=} 0
        + I(\dv_{\Ic \setminus \Bc}; \qv_{\Ic \setminus \Bc} \vert \dv_\Bc, \pv_\Bc, F_{[\Nsf]}) \stackrel{\rm{(e)}}{=} 0 %
\end{align*}
where the equality follows because:
(a) $X$ in~\eqref{eq:LayerMDSmessages:X} is a function of $(F_{[\Nsf]}, \qv_\Ic)$;
(b) $Z_\Bc$ in~\eqref{eq:LMDC:Zjall} is function of $(F_{[\Nsf]}, \pv_\Bc)$; 
(c) since $\qv_{\Bc} = \dv_{\Bc}+\pv_{\Bc}$ and thus can be dropped; 
(d) the user demands are independent, and independent of the local randomness; and
(e) $(\dv_\Bc, \pv_\Bc,\dv_{\Ic \setminus \Bc}, \qv_{\Ic \setminus \Bc})$ are independent.

\section{Example for $(\Ksf, \Ksf^\prime, \Nsf) = (6, 3, 3)$}
We provide an example to illustrate our new proposed schemes for the $(\Ksf, \Ksf^\prime, \Nsf) = (6,3,3)$ hotplug model with demand privacy against colluding users. %

\subsection{Uncoded Scheme} 
Each user caches the first $\Bsf\Msf/\Nsf$ symbols of each file, and the server sends the remaining $\Bsf(1-\Msf/\Nsf)$ symbols of each file regardless of the demands; the attained tradeoff is $\Rsf^\text{\rm(uncoded)}(\Msf)=\Nsf-\Msf$; this scheme is clearly private and the users can compute locally any functions, not just scalar linear combinations of the files. Thus, we attain the trivial corner points $(\Msf,\Rsf)=(0,\Nsf)=(0, 3)$ and $(\Msf,\Rsf)=(\Nsf,0)=(3, 0)$.
This holds for any set of parameters.

\subsection{New Scheme 1: Case $t=0$}
For $t=0$ there is no file split and no MDS coding. user~$j\in[6]$ only caches the privacy key $T(\pv_j,1)= \sum_{n\in[3]} p_{j,n} F_n$ where $\pv_j=(p_{j,1},p_{j,2},p_{j,3})$ has i.i.d. uniformly distributed entries. For active user set $\Ic=\{a,b,c\}$, the server sends 
\begin{align*}
X_{\{a,b\}} &= T(\qv_a,1) + T(\qv_b,1), \\
X_{\{a,c\}} &= T(\qv_a,1) + T(\qv_c,1),
\end{align*}
and the users locally compute $X_{\{b,c\}} = X_{\{a,b\}}-X_{\{a,b\}}=T(\qv_b,1)-T(\qv_c,1)$;
thus each user has two linear combinations of two files and can actually recover both files, which is the whole library and thus the delivery is private. The cache size is $\Msf=1$ and the load is $\Rsf=2$. 
Notice that some multicast messages need not be sent as they can be locally obtained as a linear combination of those that have already been sent.

\subsection{New Scheme 1: Case $t=1$}
\paragraph{File Partition} Partition each file as
\begin{align*}
    F_n &=( \Tbp(\ev_n,\ev_{\{1\}}), \Tbp(\ev_n,\ev_{\{2\}}), \Tbp(\ev_n,\ev_{\{3\}}) ),  \forall n \in [3].
\end{align*}

\paragraph{Cache Placement}
Let $\Gm = [\gv_{\{\ell\}}: \ell \in \Omega_{[6]}^1],$ be the generator matrix of an MDS code of rate $3/6$ , i.e., any $3$ columns of $\Gm $ are linear independent. 
Generate i.i.d. uniformly at random $\pv_j \in \mathbb{F}_\qsf^3,$ for each user~$j \in [6]$; 
since $\eta=[\binom{\Ksf^\prime}{t} > \binom{\Ksf-1}{t-1}]^+=[3-1]^+=2$, two privacy keys must be stored
\begin{align*}
    Z_j = \{\pv_j\} \cup \{\Tbp(\ev_n, \gv_{\{j\}}): n \in [3]\} \cup \{\Tbp(\pv_j, \xi_{j,\ell}): \ell \in [2]\},
\end{align*}
where $\{\xi_{j,1}, \xi_{j,2}\}$ are two columns of $\Gm$ different from $\gv_{\{j\}}$. %

We now show how users can compute the privacy keys that are not cached locally. User~$1$ knows $\Tbp(\ev_n,\gv_1)$ for $n\in[3]$, and can compute $\Tbp(\pv_1,\gv_1)  = \sum_{n\in[3]} p_{1,n} \Tbp(\ev_n,\gv_1)$. With $\Tbp(\pv_1,\gv_1)$ and the cached privacy keys, it can compute  $\Tbp(\pv_1,\gv_u)=  a_{u,1} \Tbp(\pv_1,\gv_1) + a_{u,2} \Tbp(\pv_1,\xi_{1,1}) + a_{u,3} \Tbp(\pv_1,\xi_{1,2})$  where 
 $\gv_u = a_{u,1} \gv_1 + a_{u,2} \xi_{1,1} + a_{u,3} \xi_{1,2}$  for all $u\in[6]$
since $\{\gv_1, \xi_{1,1}, \xi_{1,2}\}$ are linearly independent by the MDS property.
Similarly for the other users.

\paragraph{Delivery Phase}
As an example, for active user set $\Ic = \{1,2,4\}$ and demands  $\dv_\Ic = [\ev_1;\ev_2;\ev_3]$,
the server sends the active set $\Ic$, $\qv_\Ic = \dv_\Ic + \pv_\Ic$, and %
\begin{align*}
    X_{\{1,2\}} &= \Tbp(\ev_2+\pv_2, \gv_1) + \Tbp(\ev_1+\pv_1, \gv_2), \\
    X_{\{1,4\}} &= \Tbp(\ev_3+\pv_4, \gv_1) + \Tbp(\ev_1+\pv_1, \gv_4), \\
    X_{\{2,4\}} &= \Tbp(\ev_3+\pv_4, \gv_2) + \Tbp(\ev_2+\pv_2, \gv_4).
\end{align*}

We now show each how user can correctly decode its desired subfiles. User~$1$ misses $\Tbp(\ev_1, \gv_2)$ (in $X_{\{1,2\}}$) and $\Tbp(\ev_1, \gv_4)$ (in $X_{\{1,4\}}$). Since it has all bilinear combinations of the form $\Tbp(\pv_1, \cdot)$ and $\Tbp(\cdot,\gv_1)$, it can recover $\Tbp(\ev_1, \gv_2)$ %
and $\Tbp(\ev_1, \gv_4)$. %
Finally, with $\Tbp(\ev_1, \gv_j) : j\in\{1,2,4\}$, %
it can compute $F_1$ since $\{\gv_1, \gv_2, \gv_4\}$ are linearly independent by the MDS property.
Similarly for the other users.

Thus $(\Msf,\Rsf) = (5/3, 3/3)$ is achievable.

\subsection{New Scheme 1: Case $t=2$}
\paragraph{File Partition} Partition the files into 3 parts as
\begin{align*}
    F_n &=( \Tbp(\ev_n,\ev_{\{1,2\}}), \Tbp(\ev_n, \ev_{\{1,3\}}), \Tbp(\ev_n, \ev_{\{2,3\}})), \ \forall n \in [3].
\end{align*}
\paragraph{Cache Placement} The MDS generator matrix is $\Gm = [\gv_{\Tc}: \Tc \in \Omega_{[6]}^2]$ of rate $3/15$, i.e., any $3$ columns of $\Gm$ are linear independent. We i.i.d uniformly generate $\pv_j \in \mathbb{F}_{\qsf}^{3}$ for each user~$j \in [6]$. Since $\eta = [ \binom{\Ksf^\prime}{t} - \binom{\Ksf-1}{t-1}]^+ = [3-5]^+=0$, there is no need to store privacy keys. Then the placement is
\begin{align*}
    Z_j = \{ \pv_j \} \cup \{\Tbp(\ev_n, \gv_{\{j,u\}}): n \in [3],  u\in[6]\setminus\{j\} \}.
\end{align*}

Users compute the all privacy keys of the form $\Tbp(\pv_j,\gv_\Tc)$ similarly to how described for the case $t=1$. 

\paragraph{Delivery Phase}
For example, for active user set $\Ic = \{1,2,4\}$ and the demands $\dv_\Ic = [\ev_1; \ev_2; \ev_3]$, the server sends the active set $\Ic$, $\qv_\Ic = \dv_\Ic + \pv_\Ic$, and $X_{\{1,2,4\}}$ given by
$
    \Tbp(\ev_1+\pv_1,\gv_{\{2,4\}}) + \Tbp(\ev_2+\pv_2,\gv_{\{1,4\}})  + \Tbp(\ev_3+\pv_4,\gv_{\{1,2\}}).
$
Users decode their desired subfiles similarly to the case $t=1$. 
Thus $(\Msf,\Rsf) = (15/3,1/3)$ is achievable. Here $5 = \Msf > \Nsf = 3$, i.e., each user caches more than the whole library. The point was to show a case where users need not store privacy keys.

\begin{figure}
    \centering
    \includegraphics[width=0.4\textwidth]{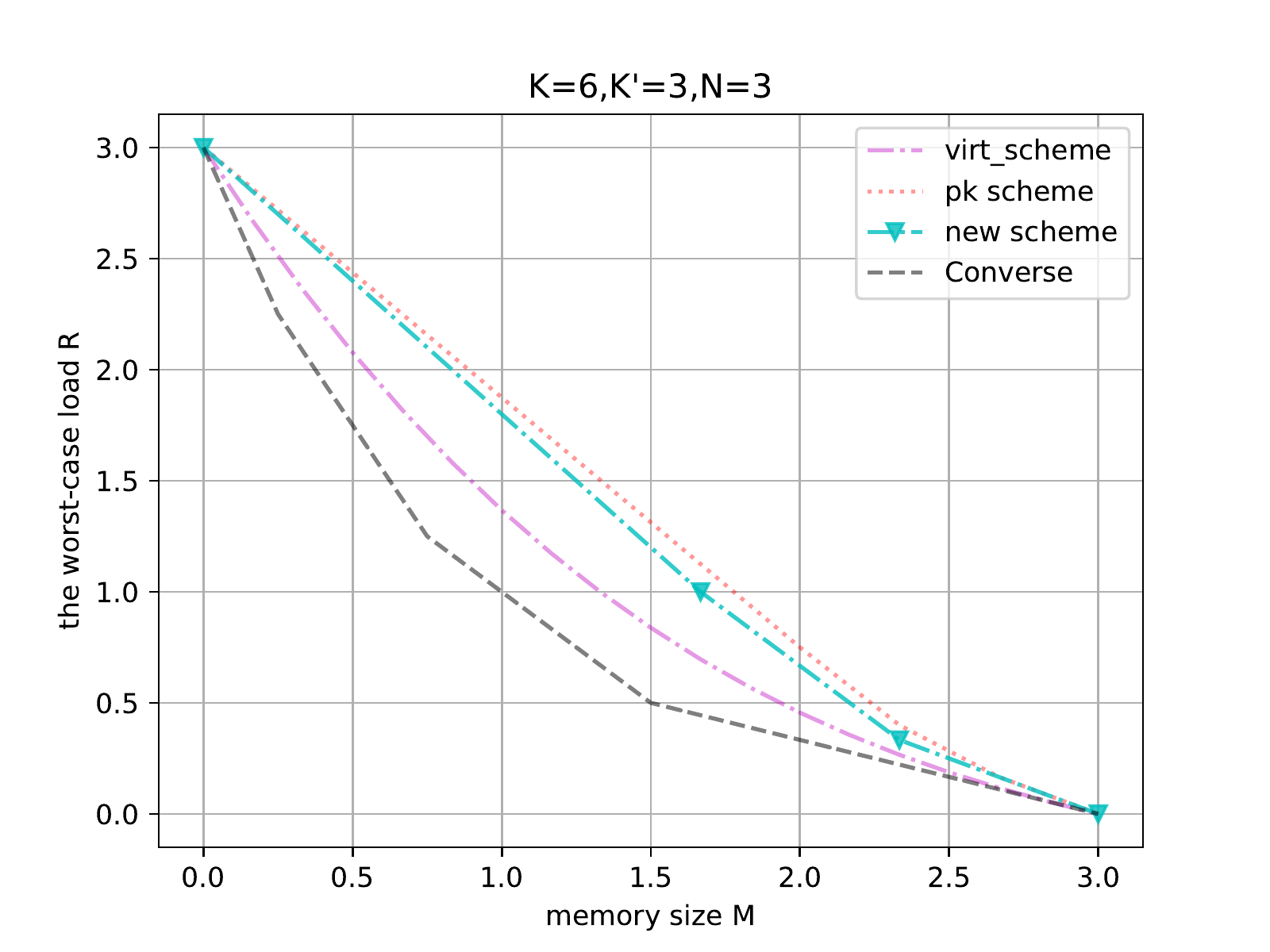} 
    \caption{\small Memory-load tradeoffs for the hotplug model with private demands when $(\Ksf,\Ksf^\prime,\Nsf) = (6, 3, 3)$.}
    \label{fig: memory-tradeoff for (6,3,3)}
\end{figure}

\subsection{Overall Performance}
Fig.~\ref{fig: memory-tradeoff for (6,3,3)} compares the performance of the baseline schemes and the new schemes (i.e., the line labeled `new scheme' is the convex hull of the corner points from Theorems~\ref{thm:layerscheme} and~\ref{thm:barescheme}). For example, the corner point $(7/3,1/3)$ is from the Theorem~\ref{thm:barescheme}.
The converse bound is derived as detailed in Remark~\ref{rem:converse}.

This is a case where a baseline scheme outperforms our new schemes; in Appendix we show that this is not the case for larger values of the parameters. We chose this case as it is the smallest example where we can clearly and simply show all the new aspects of the proposed scheme in Theorem~\ref{thm:layerscheme}.

\section{Conclusion}\label{sec:conclusion}
This paper considered the hotplug model with demand privacy. 
New achievable schemes that combine ideas from privacy keys and MDS coding are proposed. 
The performance of the new schemes is better in general than baseline schemes, and attain lower subpacketization levels. 
Open questions remain to answer, including how to improve on the baseline schemes in the mid-memory regime, and having a random number of active users.

\bibliographystyle{ieeetr}
\bibliography{references}

\appendices

\section{Proof of Theorem~\ref{thm:barescheme}}\label{apx:proofbarescheme}
\paragraph{File Partition}
We partition each file into $\Ksf^\prime$ equal-length subfiles as
\begin{align}
    F_n = (F_{n,\ell}: \ell \in [\Ksf^\prime]), \ \forall n \in [\Nsf].
\end{align}
With a notation similar to~\eqref{eq:bilinearproduct}, we let $F_{n,\ell} = \Tbp(\ev_n, \ev_\ell), n \in [\Nsf], \ell \in [\Ksf^\prime]$.

\paragraph{Cache Placement}
For each user~$j \in [\Ksf]$, we consider the generator matrix $\Gm_j = [\gv_{j,1}, \cdots, \gv_{j,\Ksf^\prime-1}]$ of size $\Ksf^\prime \times (\Ksf^\prime-1)$ 
More conditions on $\{\Gm_1, \ldots, \Gm_\Ksf, \xi_1, \ldots, \xi_\Ksf\}$ will be needed to ensure decodability and will be listed next.
For each user~$j \in [\Ksf]$, we also generate i.i.d uniformly at randomly a vector $\pv_j \in \mathbb{F}_{\qsf}^{\Nsf}$. 
We populate the caches as
\begin{subequations}
\begin{align}
  Z_j &= \left\{\Tbp(\ev_n,\gv_{j,\ell}): n \in [\Nsf], \ell \in [\Ksf^\prime-1]\right\} 
  \label{eq:BMDC:ZjMAN}
  \\ & \cup \left\{ \Tbp(\pv_j, \xi_j)\right\} \cup \left\{\pv_j\right\}, \ \forall j \in [\Ksf],
  \label{eq:BMDC:ZjPK}
\end{align}
\end{subequations}
and the memory size is
\begin{align}
    \Msf 
    = \Nsf \frac{\Ksf^\prime-1}{\Ksf^\prime} + \frac{1}{\Ksf^\prime} 
    = 1 + \frac{\Ksf^\prime-1}{\Ksf^\prime}(\Nsf-1).
\end{align}

\paragraph{Delivery Phase} Given the set of active users $\Ic \in \Omega_{[\Ksf]}^{\Ksf^\prime}$ where user~$j \in \Ic$ demands $\dv_j$, let $\qv_j = \pv_j + \dv_j$ as in PK scheme. 
The server sends 
\begin{subequations}
    \begin{align}
      X &=  ( X_\Ic ,\qv_{\Ic}, \Ic ), \\ %
      X_\Ic &= \sum_{j \in \Ic} \Tbp(\qv_j, \phi_{\Ic,j}),
      \label{eq:BMDSmessages:X}
    \end{align}
\end{subequations}
where the vectors $\phi_{\Ic,j}$ are to be chosen so as to insure users can recover their demanded linear combination of files from $X_\Ic$.
As for the PK scheme, we rewrite $X_\Ic$ as
\begin{subequations}
\begin{align}
      X_\Ic 
      &= \underbrace{ \Tbp(\dv_v, \phi_{\Ic,v}) }_{\text{desired by user~$v$}}
       + \underbrace{ \Tbp(\pv_v, \phi_{\Ic,v}) }_{\text{from cached content in $Z_v$}}
\label{eq:BMDSmessages:Xsplit1}
    \\&+ \sum_{j \in \Ic \setminus \{v\}} \underbrace{ \Tbp(\qv_j, \phi_{\Ic,j}) }_{\text{from cached content in $Z_v$ and $\qv_{\Ic}$}}, 
    \ \forall v \in \Ic,
\label{eq:BMDSmessages:Xsplit2}
\end{align}
\label{eq:BMDSmessages:Xsplit}
\end{subequations}
where in the underlines we clarified how we intend to remove from $X_\Ic$ the `interfering' terms for any user~$v \in \Ic$. We will describe next how to remove the `interfering' terms, thus also identifying additional properties we require of the generator matrices; assuming that it is indeed possible,
the load is 
\begin{align}
\Rsf = 1/\Ksf^\prime. 
\end{align}

\paragraph{Proof of Privacy} Privacy can be proved similarly to Theorem~\ref{sec:prooflayerscheme}.

\paragraph{Conditions for Decodability}

$\bullet$ Condition~1: $[\Gm_v; \xi_v]$ is full rank for all $v\in[\Ksf]$. 

Reason: The second term in~\eqref{eq:BMDSmessages:Xsplit1} can be computed from the cache content $Z_v$ as follows:  if $[\Gm_v; \xi_v]$ is full rank, then any $\tilde{\av}\in \mathbb{F}_\qsf^{\Ksf^\prime}$ can be obtained as 
\begin{align}
\tilde{\av} &= \sum_{\ell\in[\Ksf^\prime-1]} a_\ell\gv_{j,\ell} + a_{\Ksf^\prime}\xi_j, 
\end{align}
for some $\av=[a_1,\ldots,a_{\Ksf^\prime}]$, %
thus user~$v$ can compute
\begin{multline}
\Tbp(\pv_v, \tilde{\av}) = \sum_{\ell\in[\Ksf^\prime-1]} a_\ell \left(\sum_{n\in[\Nsf]} p_{v,n} \Tbp(\ev_n,\gv_{j,\ell})\right) \\ 
+ a_{\Ksf^\prime} \Tbp(\pv_v, \xi_v),
\end{multline}
for any $\tilde{\av}$, and thus in particular for $\tilde{\av}=\phi_{\Ic,v}$.

$\bullet$ Condition~2: $[\Gm_v; \phi_{\Ic,v}]$ is full rank for all $v\in[\Ksf]$ and all $\Ic\in\Omega_{[\Ksf]}^{\Ksf^\prime}$.
 
Reason:  Let us assume for now that the term in~\eqref{eq:BMDSmessages:Xsplit2} can also be computed from the cache content $Z_v$, thus user~$v\in\Ic$ has recovered $\Tbp(\dv_v, \phi_{\Ic,v})$. User~$v\in\Ic$ can compute 
\begin{align}
\Tbp(\dv_v, \gv_{v,\ell}) &= \sum_{n\in[\Nsf]} d_{v,n} \Tbp(\ev_n,\gv_{v,\ell}), \ \ell\in[\Ksf^\prime-1]
\label{eqQ:whatUcancompute}
\end{align}
but does not have the `last piece' of the demanded linear combination $\Tbp(\dv_v, \xi_v)$. 
Note that having also $\Tbp(\dv_v, \xi_v)$ would allow user~$v$ to generate any $\Tbp(\dv_v, \cdot)$, and thus recover the demanded linear combination, since $[\Gm_v; \xi_v]$ is full rank by the condition derived above.
Thus we need $[\Gm_v; \phi_{\Ic,v}]$ to be full rank as well, in order for the recovered $\Tbp(\dv_v, \phi_{\Ic,v})$ to be the equivalent to $\Tbp(\dv_v, \xi_v)$.

$\bullet$ Condition~3: For $i, j \in [\Ksf], i \neq j$, $[\Gm_i,\Gm_j]$ is a full rank matrix.

Reason: We finally need to show under which conditions the term in~\eqref{eq:BMDSmessages:Xsplit2} can be computed from the cache content of user~$v\in\Ic$. We saw that user~$v\in\Ic$ can compute terms as in~\eqref{eqQ:whatUcancompute}.
Assume the subpacketization to be $\Lsf=\Ksf^\prime$ so that each subfile is a scalar in $\mathbb{F}_\qsf$.
If we let $\Fm \in \mathbb{F}_\qsf^{ \Nsf \times \Ksf^\prime }$ be the matrix with all the subfiles, then the cached content in~\eqref{eq:BMDC:ZjMAN} is equivalent to the matrix $\Fm\Gm_j\in \mathbb{F}_\qsf^{ \Nsf \times \Ksf^\prime-1 }$. The terms in~\eqref{eq:BMDSmessages:Xsplit2} can also be expressed in terms of $\Fm$ as 
$\Tbp(\qv_j, \phi_{\Ic,j}) = \qv_j^T \Fm \phi_{\Ic,j}$ and should be computable from the cache of user  user~$v\in\Ic$, that is, there should exist a vector $\xv_{v,j}$ of length $\Ksf^\prime-1$ such that 
\begin{align}
\phi_{\Ic,j} = \Gm_v \xv_{v,j}, \ \forall j,v\in\Ic, v\not=j.
\label{eqQ:whatUmust be able to compute}
\end{align}

As an example consider $\Ic = [\Ksf^\prime]$ %
and rewrite the condition in~\eqref{eqQ:whatUmust be able to compute} for $j=\Ksf^\prime$  as 
\begin{align}
    \underbrace{
    \begin{bmatrix}
        \Gm_1   & \Gm_2   &        &  \\
        \vdots   &         & \ddots &  \\
        \Gm_1   &         &        & \Gm_{\Ksf^\prime-1} \\
    \end{bmatrix}
    }_{= \widetilde{\Gm}_{j} \in \mathbb{F}_\qsf^{ \Ksf^\prime (\Ksf^\prime-2) \times (\Ksf^\prime-1)^2 } }
    \underbrace{
    \begin{bmatrix}
        -\xv_{1,j} \\ \vdots \\ \xv_{\Ksf^\prime-1,j} 
    \end{bmatrix}
    }_{= \widetilde{\xv}_{j} }
    = 0,  \ j=\Ksf^\prime,
\end{align}
where $\widetilde{\Gm}_{j}$ is a `fat matrix', i.e., $\Ksf^\prime (\Ksf^\prime-2) < (\Ksf^\prime-1)^2$;
to ensure a non-zero solution for $\widetilde{\xv}_{j}$ we need $\widetilde{\Gm}_{j}$ to be full rank.
For~\eqref{eqQ:whatUmust be able to compute} to hold, it suffices that any matrix $[\Gm_j : j\in \Ic]$ is an MDS matrix for all  $\Ic\in\Omega_{[\Ksf]}^{\Ksf^\prime}$.

$\bullet$ For sufficiently large enough finite field size $\qsf$, if $[\Gm_1,\ldots,\Gm_{\Ksf}]$ is an MDS code, then the above three conditions are satisfied. See also an example next.

\paragraph{Example}
Assume $\Ksf^\prime = 3, \Ic = \{1,2,3\}$, and let
\begin{align}
    \Gm_1^T &= \begin{bmatrix}
        1 & 1 & 1 \\
        1 & 2 & 4
    \end{bmatrix}, \ \xi_1 = [1,3,9]^T, \\
    \Gm_2^T &= \begin{bmatrix}
        1 & 3 & 9 \\
        1 & 4 & 16
    \end{bmatrix}, \ \xi_2 = [1,1,1]^T, \\
    \Gm_3^T &= \begin{bmatrix}
        1 & 5 & 25 \\
        1 & 6 & 36
    \end{bmatrix}, \ \xi_3 = [1,1,1]^T. 
\end{align}
We shall find $\phi_3$ such that $\phi_3 = \Gm_1^T \av = \Gm_2^T \bv$, i.e.,
\begin{align}
    \underbrace{
    \begin{bmatrix}
        \Gm_1 & \Gm_2 \\
    \end{bmatrix}
    }_{\textsf{rank}(\cdot) = 3}
    \begin{bmatrix}
        \av \\ -\bv
    \end{bmatrix} &= 0, \\ 
    \Rightarrow
    \begin{bmatrix}
        1 & 1 & 1 & 1 \\
        1 & 2 & 3 & 4 \\
        1 & 4 & 9 & 16
    \end{bmatrix}
    \begin{bmatrix}
        a_1 \\ a_2 \\ -b_1 \\ -b_2
    \end{bmatrix} &= 0, \\
    \Rightarrow
    \begin{bmatrix}
        1 & 1 & 1 \\
        1 & 2 & 3 \\
        1 & 4 & 9
    \end{bmatrix}
    \begin{bmatrix}
        a_1 \\ a_2 \\ -b_1
    \end{bmatrix} &= 
    \begin{bmatrix}
        b_2 \\ 4b_2 \\ 16b_2
    \end{bmatrix}. 
\end{align}
Note that any pair of matrices $[\Gm_i; \Gm_j]$ is full rank for $i,j \in [\Ksf], i\not=j$.
Let $b_2 = 1$, then
\begin{align}
    \begin{bmatrix}
        a_1 \\ a_2 \\ -b_1
    \end{bmatrix} =
    \begin{bmatrix}
        1 & 1 & 1 \\
        1 & 2 & 3 \\
        1 & 4 & 9
    \end{bmatrix}^{-1}
    \begin{bmatrix}
        1 \\ 4 \\ 16
    \end{bmatrix} =
    \begin{bmatrix}
        -\frac{5}{2} \\ 4 \\ -\frac{1}{2}
    \end{bmatrix}.
\end{align}
We have $\phi_3 = [\frac{3}{2}, \frac{11}{2}, \frac{19}{2}]$. 
So we find a solution for $\phi_3$, such that $\phi_3$ is in the span of $\Gm_1$ and $\Gm_2$, and also $\phi_3$ is linear independent w.r.t $\Gm_3$.

We go back to check~\eqref{eq:BMDSmessages:Xsplit}. User $3$ can compute any form of bilinear product $\Tbp(\qv_3, \cdot)$ and $\Tbp(\cdot, \phi_3)$, thus it can extract $\Tbp(\dv_3,\phi_3)$ out. As $[g_{3,1},g_{3,2},\phi_3]$ are linear independent, then $\langle \dv_3, F_{\Nsf}\rangle$ is retrievable for user~$3$.

\begin{figure}
    \centering
    \begin{subfigure}[t]{0.45\textwidth}
        \centering 
        \includegraphics[width=\textwidth]{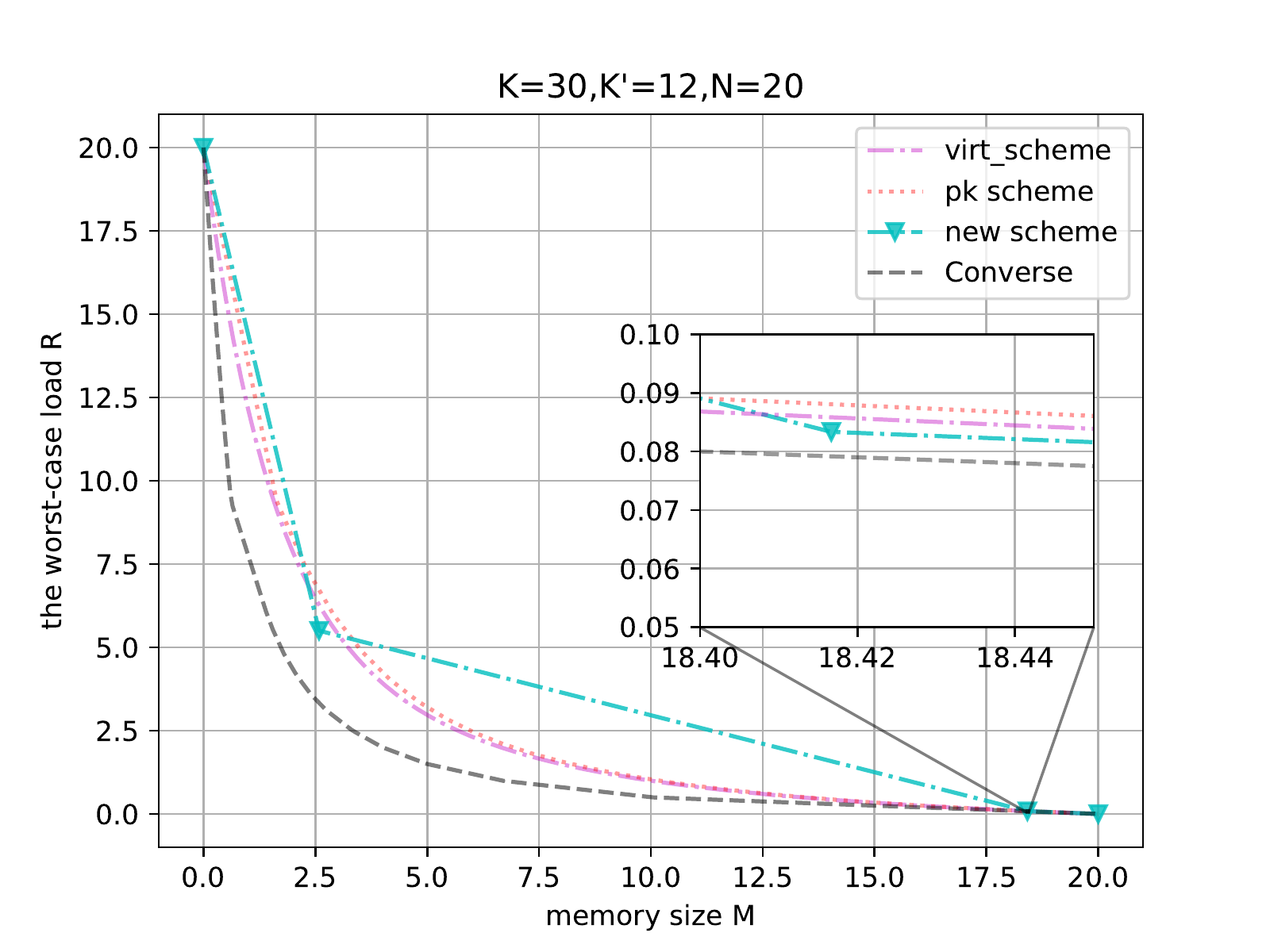} 
        \caption{Case $(\Ksf, \Ksf^\prime, \Nsf) = (30, 12, 20)$.} 
        \label{fig: memory-tradeoff for (30,12,20)}
    \end{subfigure}
    \begin{subfigure}[t]{0.45\textwidth}
        \centering 
        \includegraphics[width=\textwidth]{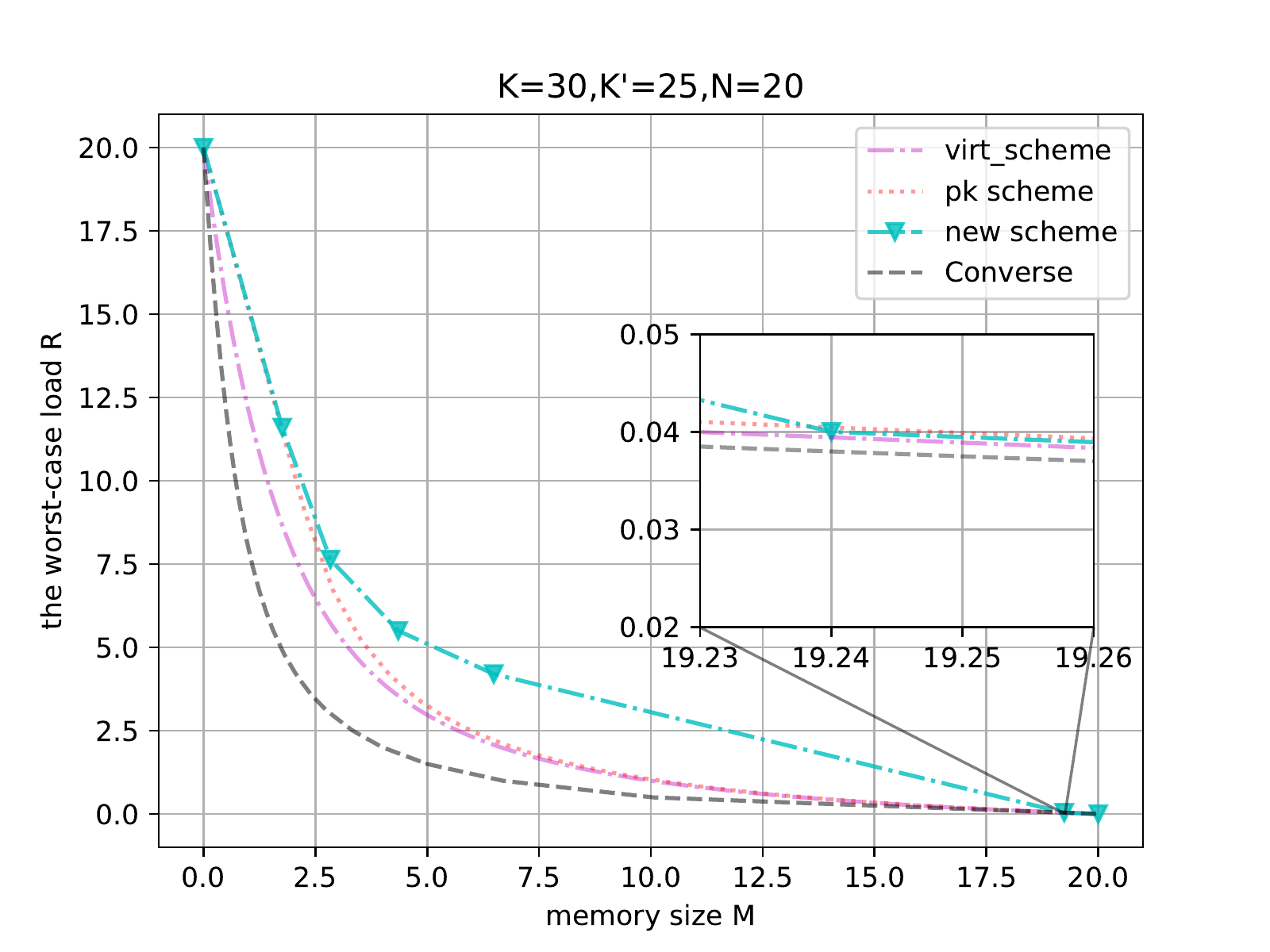}
        \caption{Case $(\Ksf, \Ksf^\prime, \Nsf) = (30, 25, 20)$.}
        \label{fig: memory-tradeoff for (30,25,20)}
    \end{subfigure}
    \caption{Memory-load tradeoffs for the hotplug model with private demands when $(\Ksf, \Nsf) = (30, 10)$ and various $\Ksf^\prime$.}
    \label{fig: memory-tradeoff for hotplug and baseline}
\end{figure}
\begin{figure}
    \centering 
    \includegraphics[width=0.45\textwidth]{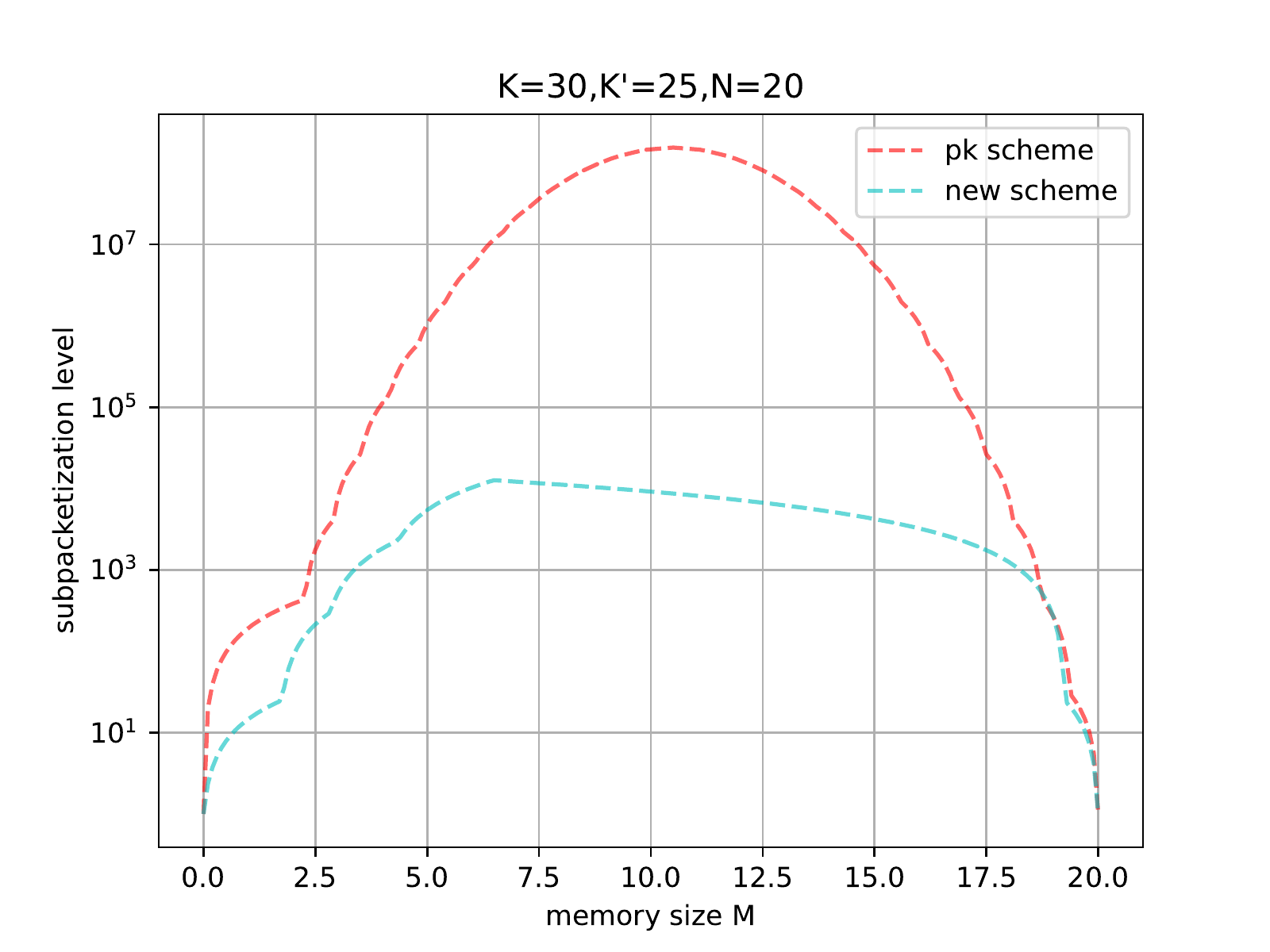}
    \caption{Subpacketization level for the hotplug system with private demands when $(\Ksf, \Ksf^\prime, \Nsf) = (30, 25, 20)$.}
    \label{fig: sublevel for (30,25,20)}
\end{figure}

\section{Numerical Evaluations}
\label{sec:numericaleval}

We provide some some examples to illustrate the performance of our new schemes, both in terms of the worst-case load and of subpacketization level. 

\indent $\bullet$ %
Fig~\ref{fig: memory-tradeoff for hotplug and baseline} shows the memory-load tradeoffs for two values of $\Ksf^\prime$ for fixed $(\Ksf, \Nsf) = (30, 20)$.
In Fig~\ref{fig: memory-tradeoff for (30,12,20)}, with $\Ksf^\prime = 12$, our schemes have good performance in both small and large memory regimes.
In Fig~\ref{fig: memory-tradeoff for (30,25,20)}, with $\Ksf^\prime = 20$, our schemes have good performance only in small memory regime.

\indent $\bullet$ Fig.~\ref{fig: sublevel for (30,25,20)} shows the subpacketization level for the case $(\Ksf, \Ksf^\prime, \Nsf) = (30, 25, 10)$. %
The subpacketization level of our new schemes is better than the privacy keys scheme; %
we did not plot the subpacketization level of the virtual users scheme since it is doubly exponential in the number of files, and thus way off the shown scale. %
\end{document}